\documentclass[%
 prd, %nofootinbib 
nofootinbib,
nobibnotes,
 amsmath,amssymb,
 aps,
]{revtex4-2}

\usepackage{graphicx}% Include figure files
\usepackage{color}
\usepackage{dcolumn}% Align table columns on decimal point
\usepackage{bm}% bold math
\usepackage[colorlinks=true,linkcolor=blue,citecolor=blue,urlcolor=blue,bookmarksopen]{hyperref}
\begin{document}

%\preprint{APS/123-QED}

\title{A Ramsey Neutron-Beam Experiment to Search for Ultralight Axion Dark Matter at the ESS}% Force line breaks with \\

\author{P. Fierlinger}
\affiliation{%
School of Natural Sciences, Technical University Munich, 85748 Garching, Germany %\textbackslash\textbackslash
}%
 %\altaffiliation[Also at ]{Physics Department, Technical University Munich}%Lines break automatically or can be forced with \\

\author{M. Holl}
\affiliation{European Spallation Source ERIC, 224 84 Lund, Sweden
}%

\author{D. Milstead}
\affiliation{%
Physics Department, Stockholm University, 
106 91 Stockholm, Sweden
}%

\author{V. Santoro}
% \altaffiliation[Also at ]{Physics Department, Technical University Munich}%Lines break automatically or can be forced with \\
%\email{valentinas.address@institution.edu}
\affiliation{
Faculty of Science, Department of Physics, Lund University, 221 00 Lund, Sweden \\
European Spallation Source ERIC, 224 84 Lund, Sweden
}%

\author{W. M. Snow}
\affiliation{%
Department of Physics, Indiana University, 
Bloomington, IN 47405, USA
}%

\author{Y. V. Stadnik}
%\email{yevgenystadnik@gmail.com}
\affiliation{School of Physics, The University of Sydney, Sydney, New South Wales 2006, Australia}

%\author{{\color{red}WHO ELSE?}}
%\affiliation{%
% ....2
% ... %\textbackslash\textbackslash
%}%

%\affiliation{%


%\affiliation{
%\affiliation{
%\affiliation{%


\date{\today}% It is always \today, today,

%%%%
\begin{abstract}
%With new neutron sources like the European Spallation Source (ESS) being available, very high intensity neutron beams offer a great opportunity to provide fundamental insights towards an improved understanding of the Universe. 
%
%In this work we discuss the sensitivity of a search for ultra-light axion-like (ALP) Dark-Matter using neutrons.
%
%The measurement is a unique opportunity to improve the  sensitivity of laboratory searches of the axion-neutron coupling parameter $C_N$ by about three orders of magnitude over a mass range of 10$^{-22}$ - 10$^{-16}$~eV.
%
%Through its coupling to the spin, the axion effectively acts like a magnetic-field on the neutron and becomes measurable via a spin precession magnetic field measurement.
%
%A possible experimental realization with a 10~m long demonstrator setup and a 50~m long extended version at a pulsed cold neutron beam in a Ramsey configuration is shown, discussing statistical and systematic limitations of such an experiment. 
%
%Both configurations are feasible to install at the HIBEAM beamline at the ESS.
%
%
%\YS{The current abstract looks a bit too long for the PRL guidelines (600 characters). I've tried to rewrite it a bit more succinctly in the following paragraph. We might need to trim a bit more.} 
%
High-intensity neutron beams, such as those available at the European Spallation Source (ESS), provide new opportunities for fundamental discoveries. 
Here we discuss a novel Ramsey neutron-beam experiment to search for ultralight axion dark matter through its coupling to neutron spins, which would cause the neutron spins to rotate about the velocity of the neutrons relative to the dark matter halo. 
We estimate that experiments at the HIBEAM beamline at the ESS can improve the sensitivity to the axion-neutron coupling compared to the current best laboratory limits by up to $2-3$ orders of magnitude over the axion mass range $10^{-22} \, \textrm{eV} - 10^{-16}$\,eV. 
%Both configurations are feasible to install at the HIBEAM beamline at the ESS. 
%The use of high intensity neutron beams like the HIBEAM beamline at the ESS therefore provides new opportunities for fundamental discoveries. 

\end{abstract}

\maketitle

%\tableofcontents

%\YS{In principle, some of the more technical details could be moved to a Supplementary Material if needed.} 

%\YS{These references were mentioned in isolation at the end of the document: nd Piegsa, PRC 88, 045502 (2013) - should we cite them somewhere?} 
% covered already.

%%%%
\section{Introduction}
\label{Sec:Intro}
Astrophysical and cosmological observations indicate that about one-quarter of the total energy and five-sixths of the total matter content of the Universe is due to dark matter (DM) \cite{particle_data_group_review_2022}, the identity and microscopic properties of which remain a mystery. 
Traditional detection schemes have largely focused on searching for possible particle-like signatures of weakly interacting massive particles (WIMPs) with masses in the $\sim \textrm{GeV} - \textrm{TeV}$ range \cite{Roszkowski_2018_WIMPs_review}, whose signatures scale to the fourth power of a small interaction constant between the DM and ordinary matter. 
On the other hand, ultralight bosons with sub-eV masses and a high particle number density may produce distinctive wavelike signatures. 
One of the leading candidates for DM is the axion, which is a light pseudoscalar (spin-0, parity-odd) particle originally proposed to resolve the strong $\mathcal{CP}$ problem of quantum chromodynamics (QCD) \cite{Peccei_1977_PQ,Weinberg_1977_axion,Wilczek_1977_axion,Kim_1979_axion,Shifman_1979_axion,Dine_1981_axion,Zhitnitsky_1980_axion}. 
Besides the canonical QCD axion, more generic axion-like particles may also exist in nature \cite{Jaeckel_WISPs_review_2010,Arias_2012_WISPs}.\footnote{In the following, we do not distinguish between the canonical QCD axion and axion-like particles and simply refer to both as ``axions''.} 
Searches for axion DM have mainly focused on the axion's possible electromagnetic coupling to photons \cite{adams_axion_2022}. 

%\YS{The following paragraph/part may benefit from further refinining and differentiation from what's already written in the abstract.} 
%
Neutron-beam experiments have previously been considered for fundamental physics tests, such as neutron-antineutron oscillations, electric dipole moment searches, new-boson-mediated forces and investigations of the structure of the weak interaction; see, e.g., Refs.~\cite{baesler_constraint_2007,nesvizhevsky_neutron_2008,pokotilovski_neutron_2011,piegsa_proposed_2012,haddock_search_2018,snow_searches_2022,Neutron_beam_EDM_2019,Neutron_beam_EDM_2022,maerkisch,oldnnbar1,santoro2023hibeam}. 
With major new neutron sources, such as the European Spallation Source (ESS) \cite{ESS_2017}, Spallation Neutron Source (SNS) in the U.S.~\cite{SNS_2006} and China Spallation Neutron Source (CSNS) \cite{CSNS_2009}, providing extremely strong, pulsed neutron beams, it is imperative to maximally leverage the potential of these state-of-the-art facilities. 
%next generations of experiments also need to be optimized for this boundary condition. 
%
%\YS{For context for the broader readership, maybe discuss recent developments in the field of neutron sources, such as the fact that the recently built/commisioned ESS provides / will provide the most intense neutron beam in the world?} 
%\YS{It looks like we currently don't cite any papers on $n-\bar{n}$ or weak interaction studies. Could someone please suggest some papers (or reviews) to cite here?} 
% Added neutron decay, old and proposed nnbar
Here we explore the sensitivity of a Ramsey experiment at such a pulsed neutron beam facility to a pseudo-magnetic field effect stemming from the coupling of an ultralight axion DM field to the neutron. 
The derivative coupling of axion DM to neutron spins would cause neutron spins to precess about the velocity of the neutrons relative to the DM halo. 
%
%Depending on the time-dependent orientation of the magnetic field of the experiment, this would cause a phase accumulated by neutrons in the beam along the flight path. 
%\YS{Suggested rewording for previous sentence: 
%
Depending on the orientation of the pseudo-magnetic field with respect to the experiment, this would cause neutrons in the beam to accumulate an additional phase as they travel along their flight path.
%} 
We estimate that with the proposed HIBEAM neutron beamline \cite{santoro2023hibeam} at the ESS, the sensitivity to the axion-neutron coupling can be improved compared to the current best laboratory limits by up to $2-3$ orders of magnitude over the axion mass range $10^{-22} \, \textrm{eV} - 10^{-16}$\,eV.

%
%%%%
\section{Theory}
\label{Sec:Theory}
Ultra-low-mass axions with very small kinetic energies can be produced efficiently via nonthermal production mechanisms, such as vacuum misalignment \cite{Preskill_1983_axion_cosmo,Abbott_1983_axion_cosmo,Dine_1983_axion_cosmo} shortly after the big bang, and can subsequently form a coherently oscillating classical field: $a(t) \approx a_0 \cos (\omega t)$, with the angular frequency of oscillation given by $\omega \approx m_a c^2 / \hbar$, where $m_a$ is the axion mass, $\hbar$ is the reduced Planck constant, and $c$ is the speed of light in vacuum. 
The oscillating axion field carries the energy density $\rho_a \approx m_a^2 a_0^2 / 2$ and behaves like a cold, nearly pressureless fluid on length scales greater than the de Broglie wavelength of the field.\footnote{Unless indicated otherwise, we adopt the natural units $\hbar = c = 1$ in this paper.} 
Cosmological and astrophysical observations rule out the possibility of axions with masses $m_a \lesssim 10^{-21}$\,eV comprising the dominant fraction of the DM \cite{Rogers_2021_Lyman,Marsh_2017_Lyman,Viel_2017_Lyman,Sibiryakov_2018_galactic}, although such ultralight axions may still comprise a sub-dominant fraction of the DM depending on their mass. 
Gravitational interactions between axion DM and ordinary matter during galactic structure formation subsequently virialise galactic axions ($v_\textrm{vir} \sim 300$\,km/s locally), which gives an oscillating axion field in our local Galactic region the finite coherence time: $\tau_\textrm{coh} \sim 2 \pi / (m_a v_\textrm{vir}^2) \sim 2 \pi \times 10^6 / m_a$; i.e., $\Delta \omega / \omega \sim 10^{-6}$ corresponding to nearly monochromatic oscillations of the field.

An axion field may interact with nucleons via the derivative coupling: 
\begin{equation}
\label{axion-nucleon_coupling}
\mathcal{L}_\textrm{int} = - \frac{C_N}{2 f_a} \partial_\mu a \, \bar{N} \gamma^\mu \gamma^5 N \, , 
\end{equation}
where $N$ and $\bar{N}$ denote the nucleon field %($N = n$ in the case of the neutron and $N = p$ in the case of the proton) 
and its Dirac adjoint, respectively, $f_a$ is the axion decay constant, and $C_N$ is a model-dependent dimensionless parameter. 
The spatial components of the derivative coupling of an oscillating axion DM field in the laboratory frame of reference, $a (t,\boldsymbol{r}) \approx a_0 \cos(m_a t - \boldsymbol{p}_a \cdot \boldsymbol{r})$, with spin-polarised nucleons in Eq.~(\ref{axion-nucleon_coupling}) simplifies as follows in the non-relativistic limit \cite{Flambaum_Patras_2013,stadnik_axion-induced_2014}: 
\begin{equation}
\label{axion-wind_Hamiltonian}
H_\textrm{int}(t) \approx \frac{C_N a_0}{2 f_a} \sin(m_a t) \, \boldsymbol{\sigma}_N \cdot \boldsymbol{p}_a \equiv \boldsymbol{\sigma}_N \cdot \boldsymbol{B}_\textrm{eff} (t) \, , 
\end{equation}
which resembles the interaction of a nucleon spin $\boldsymbol{\sigma}_N$ (which has unity norm, $|\boldsymbol{\sigma}_N| = 1$) with a time-varying \textit{pseudo}-magnetic field $\boldsymbol{B}_\textrm{eff} (t)$. 
%\footnote{Compare with the usual magnetic interaction of a magnetic dipole moment with an ordinary magnetic field, where the magnetic moment generally includes contributions from both intrinsic spin angular momenta and orbital angular momenta.} 
The interaction in Eq.~(\ref{axion-wind_Hamiltonian}) causes nucleon spins to precess about the direction of the axion DM momentum, $\boldsymbol{p}_a$, taken in the laboratory frame. 
The correlation $\boldsymbol{\sigma}_N \cdot \boldsymbol{p}_a$ in Eq.~(\ref{axion-wind_Hamiltonian}) is modulated at the daily sidereal frequency due to the rotation of Earth and can be calculated by transforming to a nonrotating celestial coordinate system; see, e.g., \cite{Stadnik_2017_book-thesis,abel_search_2017,smorra_direct_2019} for more details. 
The average value of $\boldsymbol{p}_a$, sampled over many coherence times, is expected to be directed opposite to the laboratory's orbital motion about the Galactic Center. 
%direction of motion of the laboratory relative to the DM halo. 
However, due to the stochastic nature of the axion DM field \cite{Derevianko_2016_stochastic}, the magnitude and direction of $\boldsymbol{p}_a$ during the course of measurements may differ from the long-term average value \cite{Centers_2021_stochastic,Lisanti_2021_stochastic}. 
%\YS{Consider discussing/mentioning effects due to stochastic fluctuations in DM momentum?} 

The ``axion wind'' spin-precession effect described by Eq.~(\ref{axion-wind_Hamiltonian}) induces temporal variations in the Larmor precession frequency of a polarised nucleon spin according to: 
\begin{equation}
\label{Larmor_frequency_total}
\hbar \omega_L (t) = |- \gamma_N \, \boldsymbol{\sigma}_N \cdot \boldsymbol{B} + 2 \boldsymbol{\sigma}_N \cdot \boldsymbol{B}_\textrm{eff} (t)| \, , 
\end{equation}
principally from the component of $\boldsymbol{B}_\textrm{eff}(t)$ directed along the applied magnetic field $\boldsymbol{B}$, with $\gamma_N$ being the gyromagnetic ratio of the nucleon. 
%\YS{The dominant effect at the low frequencies of main interest in this proposal arises from the component of $\boldsymbol{B}_\textrm{eff}(t)$ directed along the applied magnetic field $\boldsymbol{B}$.} 
%{\color{blue}I think the polarization is precessing, and therefore a coupling to the DM happens in 2 directions: if aligned with B, it is either faster or slower precessing, so we gain a factor 2, right?. If the pseudo magnetic rotation is around a vector in the precession plane, there would be a dressing effect at higher frequency, and no effect at very low frequency.} 
A (pseudo) magnetic-field measurement using a neutron spin is based on measurement of the phase $\phi (T) = \int_0^T \omega_L (t) \, dt$, where $T$ is the spin precession time, 
%$\phi = \omega_L t $, 
resulting in a measurement sensitivity that depends not only on the coupling constant $C_n / f_a$, but also on the axion mass: for the most sensitive scenario in an experiment when $m_a \ll \omega_L$, in which case $\boldsymbol{B}_\textrm{eff}$ is approximately constant during the free precession phase and a phase can continuously build up. 
Note that the axion-DM-induced change in the Larmor precession frequency scales linearly in the small interaction constant, in contrast to the quadratic scaling in searches for virtual-axion-mediated spin-dependent forces \cite{vasilakis_limits_2009}.

\section{\label{sec:exp}EXPERIMENTAL APPARATUS}
To measure a change in $\omega_L$, Ramsey's method of separated oscillating fields \cite{Ramsey_method_1950} is used. The method is based on the comparison of $\omega_L$ to an external frequency $\omega_1$ from a stable clock. 
%
%The method works the following way: 
Polarized neutrons enter a spin-flipper coil, a region with a magnetic field $\boldsymbol{B}_1$, oscillating at a frequency $\omega_1$. 
%\YS{The use of the symbol $\omega$ here may be confusing, as we already used it earlier to define the axion Compton frequency, and may suggest that the experiment is done ``on resonance''. Maybe we could use a different symbol like $\omega_1$ here?} 
This closely approximates a rotating magnetic field at frequency $\omega_L$ determined by the constant background field $\boldsymbol{B}_0$. 
%
%The amplitude of 
$B_1$ is adjusted to 
%the length of the coil and the velocity of the neutrons $v$, to 
ultimately result in a spin rotation by $\pi/2$ radians, into the plane normal to $\boldsymbol{B}_0$, resulting in a precession at the Larmor frequency $\omega_L$. 
The neutrons then precess in the constant magnetic field $\boldsymbol{B}_0$ for a time $T$, where a phase $\phi = \omega_L  T$ is built up. At the end of the flight path, a second $\pi / 2$-flip is applied by passage through a second coil, driven with the same current that also drives the first coil. 
In the case when $\omega_L = \omega_1$, 
%\YS{Same question as before: can we change $\omega$ to $\omega_1$ here too?} 
both $\pi/2$-flips add up, resulting in a downward pointing polarization of the neutron. 
A mismatch of the frequencies, such that a phase $\pi$ is built up during the flight time relative to $\omega_1$, 
results in the polarization pointing upwards. 
%\YS{I'm confused by the wording of the previous sentence: Frequencies can't be mismatched by dimensionless quantities like $\pi$. Presumably, what we mean here is that a phase buildup of $\pi$ radians, due to a mismatch in $\omega_L$ and $\omega_1$, results in the polarisation pointing upwards?} 
%
%Scanning different frequencies $\omega$, i.e. p
Performing many independent Ramsey experiments with different settings for $\omega_1$ results in the well-known Ramsey fringes. 
%, with each point on the curve being an independent measurement. 
%
An unknown magnetic-field-like effect, as well as any other magnetic field applied in addition to $\boldsymbol{B}_0$, will shift the whole pattern in frequency. 
%
%shift this fringe pattern sideways. \YS{Instead of ``shift ... sideways'', maybe could say ``translate ... in frequency''?} 
Repeatedly determining the central fringe of the pattern for subsequently different $\boldsymbol{B}_0$ orientations thus reveals a slowly oscillating offset field $\boldsymbol{B}_\textrm{eff}$ or its absence within the boundaries of statistical and systematic precision. 
This can then be converted into the parameter space of coupling constant and axion mass. 
%
%As an interesting detail to this proposed experiment, the 
A constant magnetic field $\boldsymbol{B}_0$ is obtained with magnetic shielding around the field-generating coil and neutron flight path. 
However, the axion DM field is not necessarily affected by the passive magnetic shield based on a ferromagnetic alloy and a (regular) conductive shield \cite{Zolotorev_2016_shielding}. %\YS{Do we mean a superconducting shield or a regular conducting shield?} 
%  and , , since axions can couple to nucleon spins without necessarily coupling to electron spins. 
%
%The ESS provides a high flux of cold neutrons, which is particularly attractive for an experimental search for spin-rotation effects due to such a magnetic-field-like effect on the neutron's spin: 
%
A flight time of $\sim 50$\,ms for a length of $50$\,m with high statistics and a 10\,Hz repetition rate allow for a sensitive bandwidth to an oscillating field with frequency from around 3\,Hz (corresponding to an axion mass of $10^{-14}\,\textrm{eV}$) to below mHz. 
%\YS{Isn't the typical flight time for a neutron with a speed of $\sim 1$\,km/s of the order of $\sim 10 - 100$\,ms for a length of $10 - 50$\,m/s?} 
%
%

\begin{figure}
\begin{center}
\includegraphics[width=.75\textwidth]{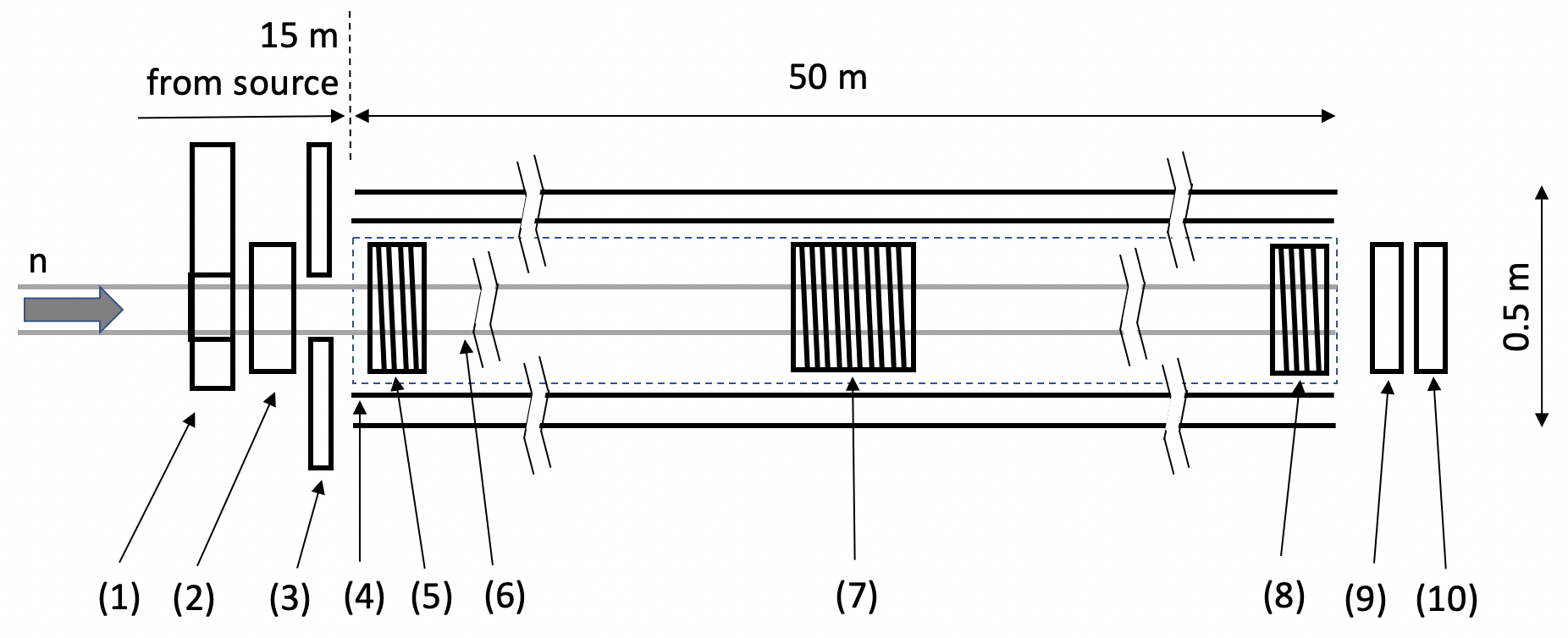}
\caption{Schematic setup of the experiment with components (longitudinal cut), neutrons entering from the left (see the main text for details). The flight path is 50\,m. 
%\YS{Since we're only showing the 50\,m case in the projections plot, do we still need to show both lengths in the figure and mention both lengths in the caption?} 
} 
\label{fig:exp}
\end{center}
\end{figure}

As shown in Fig.~\ref{fig:exp}, the apparatus to determine the spin precession frequency of the neutrons consists of: (1) a chopper; (2) a polarizer; (3) a collimator; at (4) the neutrons enter the magnetically controlled region, formed by a 2-layer passive magnetic shield, a vacuum chamber and a field coil to produce a magnetic field transverse to the neutron direction; (5) and (8) are the $\pi / 2$-spin flippers; (6) a neutron guide; %(present only for the 50~m case; 
(7) an optional $\pi$-spin flipper; (9) a spin analyzer; and (10) a detector. 
The magnetic holding field is applied over the region denoted with the dotted rectangle.
The setup is situated starting 15 meters from the moderator. 
The whole length of the flight path is magnetically shielded and a constant magnetic field is applied. 
Initial conditions for the measurement were established using the ray-tracing software McStas \cite{mcstas}. 
%
%Starting from an initial flux of 10$^{12}$ n$/$s from the ESS neutron source, 
Collimators are utilized to reduce the beam area to $10~\textrm{cm} \times 10$~cm at the entrance to the experiment. 
%\footnote{Adding a neutron guide (not part of this project) before the setup with m $=$ 1.5 would increase the intensity by a factor of 2.} before entering the magnetically shielded region. The velocity distribution has  a maximum at around 1200~m$/$s, see Fig.~\ref{initcond}.
%
%\begin{figure}
%\begin{center}
%\includegraphics[width=.9\textwidth]{init_cond.png}
%\caption{Velocity spectrum, momentum and position distribution from Mcstas simulation at 25~m distance from the ESS source.}
%\label{initcond}
%\end{center}
%\end{figure}
%
%
%
\section{\label{sec:sens}SENSITIVITY}
\begin{figure}[t]
\begin{center}
\includegraphics[width=.75\textwidth]{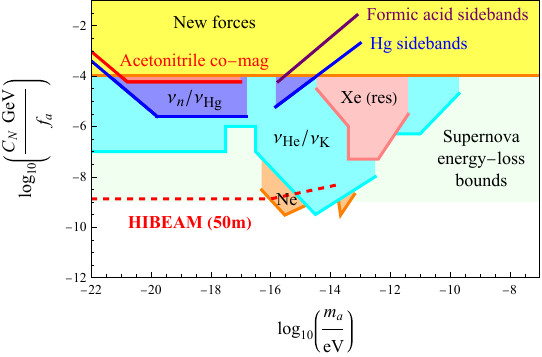}
\caption{
Projected sensitivity of a 50\,m scale Ramsey neutron-beam experiment using the HIBEAM neutron beamline at the ESS (denoted by dashed red line) to the coupling strength of axion dark matter with a neutron, defined in Eq.~(\ref{axion-nucleon_coupling}), as a function of the axion mass $m_a$, assuming one year of run time and that axions saturate the observed density of dark matter in our local Galactic region, $\rho \approx 0.4 \, \textrm{GeV/cm}^3$ \cite{particle_data_group_review_2022}. 
The cyan, blue, orange, pink, red and purple regions indicate regions of parameter space already probed by magnetometry-based searches for time-varying spin-precession effects induced by axion dark matter \cite{abel_search_2017,wu_search_2019,garcon_constraints_2019,bloch_2020_axion,jiang_search_2021,bloch_new_2022,bloch_constraints_2023,abel_search_2023,Romalis_axion_2023,Wei_2023_search,Xu_2023_constraining}. 
The yellow region denotes the region of parameter space ruled out by a magnetometry-based search for spin-dependent forces mediated by the exchange of virtual axions \cite{vasilakis_limits_2009}. 
The pale green region denotes bounds from astrophysical observations of supernovae \cite{carenza_improved_2019}, which are subject to model-dependent assumptions and may be evaded altogether \cite{Blum_2020_supernova}. 
} 
\label{fig:sens}
\end{center}
\end{figure}

{\em Statistical sensitivity.} The frequency resolution of a single Ramsey experiment scales as %$\Delta f \sim (T \alpha \sqrt{N})^{-1}$, 
\begin{equation}
\label{statist}
\Delta f = \frac{1}{2\pi T \alpha \sqrt{N}} \, , 
\end{equation}
with $N$ being the number of neutrons and $\alpha$ the contrast in detection of the different spin states. 
The sensitivity then increases with the square-root of the number of repetitions, while the oscillations of the DM field remain coherent. 
%\YS{Technically, the sensitivity to an oscillating axion DM field improves with the square root of the number of shots as long as oscillations of the DM field remain coherent, but the sensitivity only continues to grow with the fourth root of the number of shots in the temporally incoherent regime.} 
% This is what i actually meant with the bandwidth of our measurement.
The initial neutron flux from the moderator into the exit of the neutron extraction system is $1 \times 10^{12} \, n$/s, which is subsequently reduced by polarization ($\times 0.5$), polarization analysis ($\times 0.5$) and calibration runs ($\times 0.8$). 
Here we assume a quasi-perfectly fit elliptic neutron guide section with $m \sim 3$ to capture practically the whole flux. 
Also, the spin flippers are assumed to be perfectly efficient over the entire velocity range. 
%
%
%according to the values in Tab.~\ref{Tab:Flux_losses}. 
%: 
%\begin{table}
%\caption{Neutron flux estimate for a 10\,m long demonstrator setup at the HIBEAM beamline. 
%\YS{Does the neutron guiding and collimation factor include the use of a neutron guide shown in Figure \ref{fig:exp}?} 
%   The neutron guide is only assumed for the 50 m case.
%
%\YS{Maybe rename the second column as ``Reduction factor'' and remove the $\times$ signs below?} 
%} 
%\begin{ruledtabular}
%\begin{tabular}{lr}
%Contribution  & Reduction factor \\
%\hline
%Neutron guiding and collimation & $ \times $ 1 \\
%Chopper to reduce the pulse duration & $ \times $ 1 \\
%Polarization loss & $ \times $ 0.5 \\
%Limited velocity acceptance of spin flipper  & \times $ 0.2 \\
%Selection of central region of spectrum & $\times$ 0.5 \\
%Polarization analysis loss & $\times$ 0.5 \\
%Calibration runs between Ramsey cycles & $\times$ 0.8 %\\
%end{tabular}
%\end{ruledtabular}
%\label{Tab:Flux_losses}
%\end{table}
%
Assuming only systematic errors uncorrelated with the frequency of the DM oscillation signal, the experiment is statistically limited. 
A frequency resolution of $10^{-8}$~rad$/$s 
%\YS{Just to double-check, this is phase resolution as opposed to angular frequency resolution?} 
can be reached in one year of run time for a 50-meter flight path. 
As the signal is measured as a time-dependent modulation of the magnetic field, the effect alternatingly adds and subtracts from the magnetic field, thus effectively doubling the precision to $5 \times 10^{-9}$~rad$/$s. 
%demonstrator version and a factor 10 \YS{5?} improvement for the 50\,m version. 
%
Using Eq.~(\ref{axion-wind_Hamiltonian}), this translates into the sensitivity estimates shown in Fig.~\ref{fig:sens}, assuming that axions saturate the observed density of DM in our local Galactic region, $\rho \approx 0.4 \, \textrm{GeV/cm}^3$ \cite{particle_data_group_review_2022}. 
%
%\YS{I like that we get straight to the punchline here by referring to the experimental sensitivity estimates in Figure~\ref{fig:sens}. However, Fig.~\ref{fig:sens} currently appear after Fig.~\ref{fig:simparams} - should we switch the positions of these two figures?} 
%Figure~\ref{fig:sens} shows the corresponding sensitivity to axion DM from the frequency measurement with a 10-meter \YS{(curve/line description 1)} scale experiment and a 50-meter \YS{(curve/line description 2)} scale experiment, assuming 1 year of run time in both cases. 
%(dashed blue line), assuming a systematic limit of 10$^{-6}$~rad$/$s and 1 year of run time. 
%We expect up to {\color{red}four orders of magnitude} improvement in sensitivity compared to the current best laboratory limits using the proposed 10-meter scale experiment and {\color{red}a further order of magnitude} improvement using a full-scale 50-meter experiment. 
%We estimate that a future 65-meter scale experiment at the ESS would achieve an additional two orders of magnitude improvement in sensitivity. 
%
The sensitivity is also constrained by systematic issues, 
which is estimated by a simulation of the so-called Ramsey fringes, which is the polarization $P_z$ along the magnetic field $\boldsymbol{B}_0$ as a function of the detuning from the resonance frequency $\omega_L$. 
%
%Parameter variations 
The parameters varied in the simulation are the spread of measurements with different detuning, neutron statistics, duration of the neutron pulse, magnetic field maps, spin-flip pulses, neutron positions, momentum directions and velocities, 
%
%this is what is included.
%\YS{Are these the input parameters used in the simulations? It would be good to explicitly clarify what is meant here.} 
%The parameters \YS{used/varied in the simulations were?} are 
%:
%
%\begin{equation}
%\label{ramsey}
%P_z(2\tau+T) = 1- \frac{8\Omega_R^2sin^2 ( \frac{\Omega_{eff}\tau}{2} ) [ \Omega_{eff} cos (\frac{\Omega_{eff}\tau}{2})  cos (\frac{T\delta \omega}{2})       
%- \delta \omega sin(\frac{T\delta\omega}{2}) sin(\frac{\Omega_{eff}\tau}{2} )
%]^2   }{\Omega_{eff}^4} , 
%\end{equation}
%with 
the precession time $T$, the duration of a $\pi /2$-flip $\tau$, the frequency $\Omega_R$ related to $B_1$ (the Rabi frequency), % $\Omega_{eff} = \sqrt{\delta \omega^2 + \Omega_R^2}$ 
and the detuning $\delta \omega$ from the Larmor frequency. 
%The simulation draws from the distributions of input parameters generated by McStas. 
%It is employed to assess the influence of uncertainty associated with each parameter on the precision of the Ramsey experiment.
%

\begin{figure}
\begin{center}
\includegraphics[width=.75\textwidth]{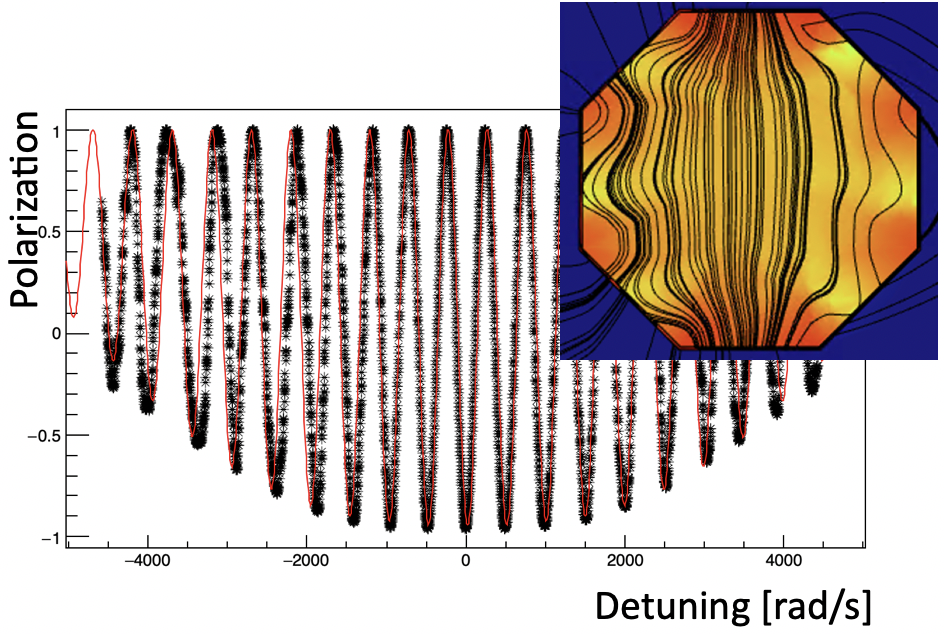}
\caption{Ramsey fringes for a realistic magnetic field. The inset shows a cut through the octagonal field setup with illustrated field lines.}
\label{fig:simparams}
\end{center}
\end{figure}

{\em Magnetic field. }
The phase buildup during free precession depends on the homogeneity of the magnetic field. 
For the magnetically controlled region, we aim for a typical homogeneity of the magnetic field over most of the volume covered by the neutron beam at the level of 10$^{-3}$, obtained with a field strength of $B_0 \sim 10 - 100$\,$\mu$T. 
Shielding of environmental disturbances is obtained by a two-layer octagonal passive magnetic shield with open ends and its axis approximately aligned with the neutron beam central axis; a coil made from 8 longitudinal wires produces a transverse magnetic field with strength $B_0 \sim 10 - 100$\,$\mu$T. 
Static residual fields smaller than 100\,nT after degaussing do not affect the measurement and the relative inhomogeneities of the magnetic field over most of the volume covered by the neutron beam are about 10$^{-3}$. 
The actual complexity of the coil configuration inside the shield and the cross-sectional area of the shielded region occupied by the neutron beam will ultimately define the homogeneity of the field experienced by the neutrons. 
The inset in Fig.~\ref{fig:simparams} shows a cut through the envisaged configuration, providing such a field with only a few wires.
%
%
%intended in a different context?} (see Fig.~\ref{fig:simparams}). %(a) shows a transverse cut through such a field together with Ramsey fringes in such a field and (b) the average magnetic field for different trajectories taken from a McStas~\cite{mcstas} simulation.
%In %Fig.\ref{fig:simparams}, the impact of the field homogeneity in the beam region is illustrated. Figures (a) and (b) represent the expectation, taken from a COMSOL
%\cite{comsol} simulation of the generated field in our shield- and coil geometry with nominally 10~$\mu$T, stray fields from the open ends and an approximately 100~mm off-centered neutron beam.  
%
%Figure~\ref{fig:simparams} illustrates the Ramsey fringes for the suggested experiment for different relative inhomogeneities and the average magnetic field for different trajectories taken from a McStas~\cite{mcstas} simulation.
% (c) is shown  a histogram of integrated magnetic fields for different neutron trajectories.
%
%
%
%. 
%we aim for a typical , obtained with a field $B_0 \sim 10 - 100$~$\mu$T.
%, such that static . 
%
%The magnetic field $B_0$ is applied transversely through an approximated "cosine-theta"-like configuration, which is achieved by a constant current through 8 wires placed longitudinally inside the shield.
%
%Shielding of time-dependent external distortions is achieved at (approximately) $<$~10~Hz through magnetizable material ("mumetal"), at higher frequencies through Eddy-current shielding in a vacuum tube surrounding the neutron beam, which is made from 8~mm thick aluminum. 
The vacuum tube, which also acts as a shield for higher frequencies, is accessed for pumping at the detector region only, as the vacuum requirement is only 10$^{-5}$~mbar. 
The passive shield design is similar to that in Ref.~\cite{wodey}. 
%\YS{What reference is this?} 
% this is a shield we built for an atomic fountain, very similar to what is needed for this experiment
%

%%%
{\em Polarization.} 
When the neutrons enter the shielded region, they pass through a cell containing polarised $^3$He. 
Through spin-dependent nuclear absorption of neutrons on $^3$He, an almost perfectly polarized neutron beam ($>$~99\%) can be obtained, thus hardly affecting contrast in the Ramsey measurement.
Polarization analysis after the Ramsey experiment can be done in the same way. 
We use cells made from GE-180 glass to obtain spin life-times of the order of a day, with polarization done using the spin-exchange optical pumping technique  with a similar setup as in \cite{hexe_EDM} with a length of several cm, a diameter of 0.1~m and a pressure of up to 10~bar. Cell designs are conceptually based on \cite{chupp3he}. The product of polarization and analysis power determines the contrast parameter $\alpha$ in Eq.~(\ref{statist}). 
Spin-flipper coils are composed of approximately 50 turns of wire wound around a nonmetallic support inside the vacuum chamber to minimize coupling to the chamber and shields; both spin-flipper coils are serially connected and fed by a function generator, stabilized by a GPS-locked frequency standard. 
While the $\pi /2$ condition is only perfectly valid for a single velocity, the analysis works for a range of velocities but with reduced contrast. 
With a $B_1$ amplitude of $\sim 100 \, \mu$T, corresponding to 3000 spin rotations per second, and for a $\pi /2$-flip at 1200\,m$/$s, the corresponding coil length is 0.1\,m. 
Placed along the flight path, a $\pi$-flipper is installed, which can be deployed. 
In particular, it can reduce wash-out of the phase measurement due to geometry; e.g., in the form of parity-symmetric magnetic-field deformations along the shield for a centered beam. 
A further interesting feature is that it can recover the flip angles for velocities that do not match the $\pi /2 $-flip conditions and thus massively broaden the velocity acceptance of the Ramsey setup. 
However, for our new physics search, a symmetric placement in a constant field also strongly suppresses sensitivity to the new-physics signal in the most interesting frequency range and thus will only be used for instrument characterization. 
It should be noted that the magnitude of $\boldsymbol{B}_0$ does not change the new physics reach. 
However, a smaller field makes the experiment more sensitive to background gradients. 
The second spin flip and the polarization analysis are similar to the first spin flip and polarizer, respectively. 
%
%The central fringe of the Ramsey experiment is fitted for analysis, and the frequency uncertainty is then

%%%
{\em Pulse duration and velocity spread.}
%
%From a Mcstas simulation, the neutron distribution is
%with their average magnetic field along their flight path. The beam is collimated to 10$ \times $10~cm at 10~m distance from the source, the divergence results from the exit of the moderator, as no neutron guide was added for this estimate.
%
%As it turns out, next to the neutron beam collimators well before the start of the magnetically controlled region, also a 
The pulse duration of a pulse at the ESS is $\sim 3$\,ms at a 10\,Hz repetition rate. 
Although the phase information is washed out, a fit of the central fringes of the Ramsey experiment can in principle resolve the phase information. However a chopper can be used to refine the pulse. 
{\em Neutron detection.} 
When $2 \times 10^{10}$ neutrons per pulse arrive within $\mathcal{O}(100)\,\textrm{ms}$ at the detector, the count rate is $\sim 10^{11}$\,s$^{-1}$. 
For a detector with 1000 pixels, the rate becomes $\sim 100$\,MHz. 
In this project, we will use a 2D-position-sensitive segmented ion chamber using $^3$He as the absorber in the gas. 
%
%. It will utilize a mixture of 3He gas and a second gas, which will be determined to optimize both the time response and the stopping distance of the ions resulting from from the n+3He $\to$ 3H + p reaction. The ion chamber will operate in current mode and with low-noise analog electronics the statistical accuracy of the detector will be dominated by the neutron shot noise in the beam. 
Similar detectors have been successfully used in several sensitive polarized neutron transmission experiments conducted at both pulsed and continuous neutron sources to search for parity-odd neutron interactions with nuclei and for possible exotic spin-dependent neutron interactions \cite{neutrondetector1}. 
%
%
%\YS{The text below appears to be a duplicate of (similar) text earlier in this section.} 
%
%{\color{red}
%The limit of 10$^{-6}$~rad$/$s can thus be reached statistically in one year for the demonstrator setup and 
%50~m ....
%
%further reduction of the chopper window by a factor %of 10 results in a systematically limited sensitivity of 10$^{-7}$~rad$/$s, or 1.5 years of run time. 
%
%
%\begin{figure}
%\begin{center}
%\includegraphics[width=.9\textwidth]{neutronflux.png}
%\caption{Loss estimates for the neutron flux.}
%\label{fig:neutronflux}
%\end{center}
%\end{figure}
%
%
%{\em Axion DM sensitivity.} 

%Figure~\ref{fig:sens} shows the corresponding sensitivity to axion DM from the frequency measurement (dashed blue line), assuming a systematic limit of 10$^{-6}$~rad$/$s and 1 year of run time. 
%With this proposed 10-meter scale experiment, we expect up to four orders of magnitude improvement in sensitivity compared to the current best laboratory limits. 
%
%
%We estimate that a future 65-meter scale experiment at the ESS would achieve an additional two orders of magnitude improvement in sensitivity. 
%}
%

%%%%
\section{Discussion and Conclusions}
\label{Sec:Conclusions}
Neutron physics offers the potential to probe new physics in ultralight axion DM searches. 
As shown in Fig.~\ref{fig:sens}, our proposed experiment at the HIBEAM neutron beamline at the ESS has significant discovery potential, offering up to $2-3$ orders of magnitude improvement in sensitivity compared to the current best laboratory limits. 
Using free neutrons as a probe of the ``axion-wind'' spin-precession effect in Eq.~(\ref{axion-wind_Hamiltonian}) provides a clean probe of the neutron interaction parameter $C_n/f_a$, which is complementary to the magnetometry approaches in Refs.~\cite{abel_search_2017,wu_search_2019,garcon_constraints_2019,bloch_2020_axion,jiang_search_2021,bloch_new_2022,bloch_constraints_2023,abel_search_2023,Romalis_axion_2023,Wei_2023_search,Xu_2023_constraining} that involve nuclei and hence probe linear combinations of proton and neutron interaction parameters \cite{stadnik_nuclear_2015}. 
Our proposed experiment is expected to be competitive with bounds from astrophysical observations of supernovae \cite{carenza_improved_2019}, whilst also offering the advantage of a much cleaner and better controlled environment. 
On the other hand, the interpretation of rare astrophysical phenomena, such as supernova explosions, which take place in conditions drastically different from those in the laboratory, require additional model-dependent assumptions that may not be valid \cite{Blum_2020_supernova}.

\begin{acknowledgments}
%{\color{red}Please add funding sources here.} 
The work of Y.~V.~S.~was supported by the Australian Research Council under the Discovery Early Career Researcher Award No.~DE210101593. 
W.~M.~S.~acknowledges support from the US National Science Foundation (NSF) grant PHY-2209481 and the Indiana University Center for Spacetime Symmetries. 
D.~M and V.~S gratefully acknowledge support from the Council for Swedish Research Infrastructure for the Swedish Research Council for the grant ‘HIBEAM pre-studies’. 
V.~S gratefully acknowledge support from the Stiftelsen för Strategisk Forskning for the grant ‘Development of a magnetic control beamline for fundamental physics and condensed matter science at the European Spallation Source’.  
%The work of Y. V. S. was supported by the Australian Research Council under
%the Discovery Early Career Researcher Award No. DE210101593.
\end{acknowledgments}

%\appendix

%\printbibliography

%\bibliographystyle{ieeetr}
\bibliography{references}

%bibliographystyle{apsrmp4-1}

\end{document}